\documentclass[amsmath,amssymb,prfluids,preprint, longbibliography]{revtex4-2}

\usepackage{float}
\usepackage{graphicx}
\usepackage{caption}
\usepackage{subcaption}
\usepackage{tabularx, booktabs}
\usepackage{xcolor}
\usepackage{hyperref}
\usepackage[normalem]{ulem}
\usepackage{txfonts}
\definecolor{yel}{HTML}{F7BA17}
\definecolor{dred}{HTML}{b01c0a}
\definecolor{dblue}{HTML}{054f87}
\definecolor{dpple}{HTML}{5e30e8}

\definecolor{lblue}{HTML}{21deff} 
\definecolor{lple}{HTML}{f50aff}

\begin{document}
\title{Creeping flows through confined arrays of cylinders}

\author{S. K. Bohling}
\affiliation{Department of Mechanical Engineering, University of California, Santa Barbara, California 93106, USA}

\author{S. S. Tanikella}
\affiliation{Department of Mechanical Engineering, University of California, Santa Barbara, California 93106, USA}

\author{J. P. Raimondi}
\affiliation{Department of Mechanical Engineering, University of California, Santa Barbara, California 93106, USA}
\affiliation{Department of Aerospace and Mechanical Engineering, University of Southern California, Los Angeles, California 90089, USA}

\author{N. D. Jones}
\affiliation{Department of Mechanical Engineering, University of California, Santa Barbara, California 93106, USA}
\affiliation{Department of Mechanical and Civil Engineering, California Institute of Technology, Pasadena, California 91125, USA}

\author{E. Dressaire}
\email[]{dressaire@ucsb.edu}
\affiliation{Department of Mechanical Engineering, University of California, Santa Barbara, California 93106, USA}
\affiliation{}

\date{June 29, 2026}

\begin{abstract}
Hair-covered appendages serve a variety of purposes in Nature, from chemical sensing and particle capture on the antennae of a crustacean to drag generation on bristled wings. At low to intermediate Reynolds numbers, these finite porous media experience three flow regimes. At low Reynolds numbers, the flow goes around the porous structure; this is the paddle or rake regime. As the Reynolds number increases, so does the relative flow rate through the array. If the fluid exits the structure mostly laterally, the flow is in the deflection regime. If the fluid exits downstream, the flow is in the sieve regime. Confining structures, such as the animal body or larger hairs, have been hypothesized to focus the flow on the hair-covered region. We investigate the influence of confinement on the flow through and around an array of cylinders, using a combination of experiments and numerical simulations. Experimentally, we vary the cylinder spacing, channel dimension, and flow rate and measure the velocity field using Particle Image Velocimetry. After comparing the results of finite element analysis with the experimental data, we numerically investigate a broader range of system geometries and flow parameters. Our results show that the confinement focuses the flow in the array and shifts the domains of existence of the three regimes. We present an analytical model that relies on the permeability of rectangular slits to predict the relative flow rate through and around the array. The model is in quantitative agreement with the numerical results, demonstrating that the flow through the array increases with increasing confinement, while the flow angle decreases. These results should provide insight into the morphology of hairy surfaces and have implications in the design of bio-inspired flow sensors and filters.
\end{abstract}


\maketitle


\section{Introduction}

Low-Reynolds-number flows through porous structures are common in both natural and engineered systems. Despite the complex interplay between the fluid and the porous structure, biological systems leverage different flow behaviors to meet their needs. For example, filter-feeding species utilize passive hairs to filter and divert food particles \cite{Riisgard2010, Koehl2001, Thiel2015, Sanderson2016, Hamann2022}. Adult barnacles \textit{Semibalanus balanoides} control the local water flow through and around short cirri by sweeping long cirri at varying speeds over different periods \cite{Thiel2015}. Comb-wing insects rely on bristled, feather-like wings to fly. Viscous effects trap air in between the bristles, and the porous structure behaves like a solid surface with a lower weight \cite{Barta2006, Jones2016, Lin2023, Kolomenskiy2025}. The reduced flow through an array of filaments is also responsible for increasing the drag on dandelion seeds, allowing them to travel long distances \cite{Cummins2018, Zhou2025}. These examples demonstrate the range of organisms, environments, and conditions in which flow can be controlled by an array of slender bodies, typically cylinders in close proximity to one another.

Technological systems seek to control and optimize the relative amount of fluid that travels through and around a finite porous medium for passive flow control, particle transport and filtration, and locomotion. Passive flow control leverages the nonlinear evolution of the flow rate entering the porous array for mixing \cite{Zhang2023} and rectifying the flow of Newtonian \cite{Alvarado2017} and non-Newtonian fluids \cite{Haward2020}. Natural filters with different structures can sort and capture particles under a variety of flow conditions and nature-inspired filters aim to extend the lifespan of traditional engineered filters. Bio-inspired filtration systems demonstrate the role of obstacles on particle separation at intermediate Reynolds numbers $Re$  \cite{Park2014, Shen2014, Cha2022}, the influence of the porosity of the filtration \cite{Sanderson2016, Yu2024}, and the leakiness of channels used in cross-flow filtration \cite{Mao2024, Hamann2025}. Millimeter-scale flying or gliding robots operate at low $Re$, much like seeds and insects. To reduce their weight and improve stability, some designs rely on bristled wings \cite{Zussman2002, Kasoju2021, Iyer2022, Sanchez2025}.

Understanding how the geometry of the porous structure and the flow properties determine the flow rates through and around the structure would therefore inform the design of a variety of bio-inspired systems. Two parameters of the porous structure are of particular interest: its porosity, i.e. void fraction and its confinement, i.e. the ratio of its width to the width of the channel. Despite extensive work on flow through and around porous objects \cite{Strong2019, Tang2019, Tamayol2011, Pezzulla2020, Hood2019}, a predictive description of the flow through a confined array of cylinders or slender bodies with high porosity remains lacking. Theoretical models, such as Brinkman's model, typically consider two flow domains with contrasting permeabilities \cite{Bear2018}. The steady incompressible flow around the object, i.e. in the high permeability region, is described by the continuity and Stokes equations at low $Re$ numbers:
\begin{equation}
    \mathbf{\nabla}.\textbf{v} = 0 \hspace{1cm} \mbox{and} \hspace{1cm}  \mu \nabla^2 \mathbf{v} - {\mathbf{\nabla}} p = 0,
\end{equation}
where $\mu$ is the fluid viscosity, and $\mathbf{v}$ and $p$ are the velocity and pressure fields respectively.
Within the object, the flow through the void space or pores is described with Darcy’s law, which relates the pressure gradient across the porous medium to the flow rate $q$, the fluid viscosity $\mu$, and the permeability $k$ of the porous medium:
\begin{equation}
    q = - \frac{\mu}{k}\nabla p 
\end{equation}
The flow descriptions in the two domains are then related through boundary conditions. This approach can be used to model the flow through simple systems such as a thin porous medium \cite{Durlofsky1987, Hwang2010}. Yet in the more complex geometry of a highly porous array of cylinders, this approach is not conducive to practical predictions, as the permeability and the boundary conditions are nontrivial. In consequence, our understanding of low- to intermediate-$Re$ flows through a finite array of cylinders is based on experimental and numerical results.

\begin{figure}
\centering\includegraphics[width=0.9\linewidth]{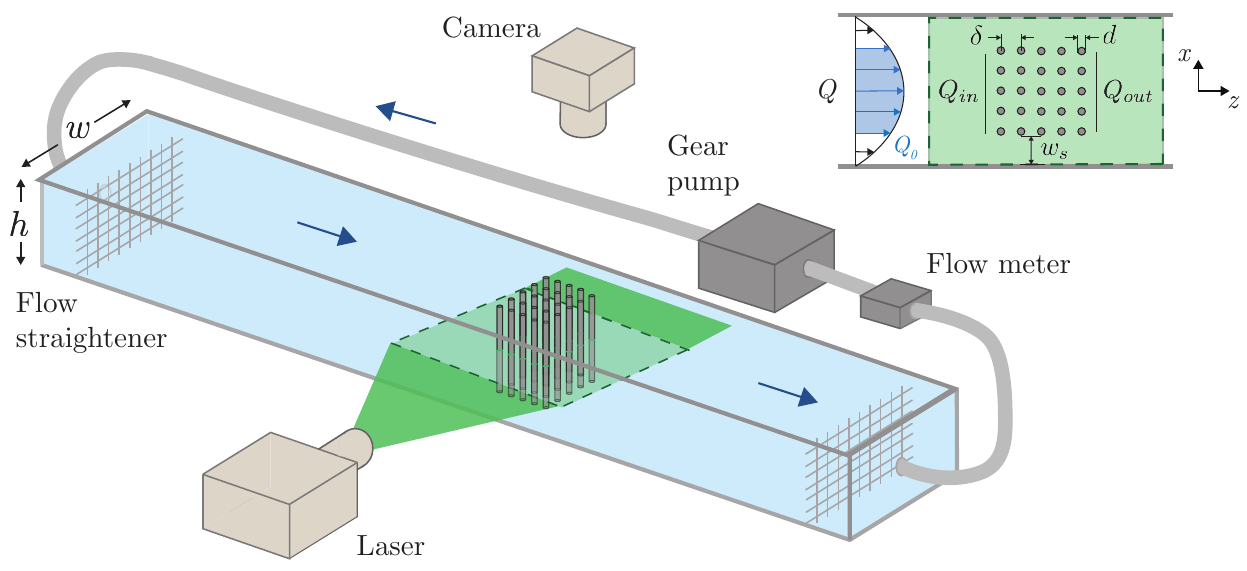}
\caption{Schematic of the experimental setup. Water of density $\rho$ and viscosity $\mu$ is pumped into a rectangular channel of width $w$, height $h = 40$ mm, and length $L= 2.1$ m (not represented to scale). A square array of $5 \times 5$ cylinders of height $h$, diameter $d$, and center-to-center spacing $\delta$ is placed $2/3$L from the inlet. The spaces between the outer cylinders of the array and the side walls of the channels are called side channels and have a width $w_s$. The flow rates in the mid-plane of the channel ($y=0$) are measured using Particle Image Velocimetry (PIV). Inset: cross-section of the channel in the plane of the laser sheet ($y=0$).}\label{fig:expsetup}
\end{figure}

Previous work on flow through and around an array of cylinders focused on  $Re = \rho d U/\mu \in[10^{-3}, 40]$, where $d$ is the diameter of the cylinders, $U$ the average velocity of the flow, and $\rho$ and $\mu$ the density and viscosity of the fluid, respectively. The existence of different flow regimes was established: the \textit{rake}, \textit{sieve}, and \textit{deflection} regimes \cite{ Cheer1987, Hood2019, Bohling2025}. In the rake regime, most of the flow is diverted around the array while the fluid in the array remains relatively stagnant. In this regime, the porous array behaves like a bluff body. The rake or paddle regime was first reported and modeled by Cheer and Koehl \cite{Cheer1987} for two infinite cylinders. Cheer and Koehl defined the leakiness of the porous structure $\Lambda$ as the ratio of the flow rate through the pores, i.e. between the cylinders $Q_{in}$, to the undisturbed flow rate $Q_0$, i.e., the flow rate in the same region, without the cylinders (see Fig. \ref{fig:expsetup}). The rake regime corresponds to low leakiness, typically defined by an order of magnitude difference between $Q_{in}$ and $Q_0$: $\Lambda <0.1$ \cite{Cheer1987, Hood2019}.
In the sieve regime, the leakiness is no longer negligible. This regime is characterized by a low curvature of the streamlines, as the flow rates entering and exiting the array are comparable. 
Finally, another regime can be observed for finite values of the leakiness, the deflection regime. Similar to the sieve regime, this flow regime is associated with a non-negligible flow rate entering the array. However, a portion of the fluid entering the array is diverted out through the side of the array and into the free stream. Curved streamlines indicate that only a fraction of the fluid entering the array exits the array downstream. This passive flow redirection could be of interest for flow control applications. 
Depending on the array porosity and the Reynolds number of the flow, a porous medium can experience one of three flow regimes \cite{Hood2019, Bohling2025}. Both experimental and numerical models predict a transition from rake to sieve and deflection as porosity and $Re$ increase. This result is consistent with the reduction in the size of the viscous boundary layer around each cylinder, allowing more fluid to enter the array \cite{Hood2019}.

In this study, we focus on wall effects on fluid transport through an array of cylinders. The wall effects are particularly important as the confinement of the array increases. 
In biological systems, confining structures, such as the animal body or larger hairs, have been hypothesized to focus the flow onto the hair-covered region \cite{Cheer1987} and to increase the leakiness. In engineered applications, confinement is common, as porous media are placed in channels that limit the flow around the array. The influence of the walls or no-slip boundary conditions on the flow through the array and on flow regimes has yet to be systematically investigated. Previous work for $Re$ numbers up to 100 has demonstrated the influence of confinement on the drag force on solid \cite {Semin2009, Wexler2013} and porous objects, such as a thin perforated plate \cite{Strong2019}. The influence of confinement on the flow within a simple porous structure was reported for a pair of confined cylinders in a microfluidic channel \cite{Zhang2023}.

In this study, we aim to quantify the effect of confinement on leakiness and develop a predictive model of the flow rates and flow regimes.
We consider square arrays of $5\times 5$ cylinders and study the flow through and around the array for different porosity and confinement conditions, as the Reynolds number of a cylinder varies from $Re = 10^{-3} $ to $ 45$. Our experimental study focuses on measuring the leakiness of confined arrays as a function of $Re$. We vary the confinement by assembling multiple flow channels; experimental measurements enable benchmarking numerical simulations. Additional numerical simulations then expand the range of porosity, confinement, and flow conditions considered, and give access to additional variables, such as the flow rate exiting the array and the pressure distribution. The results support the development of a first-order predictive model for the leakiness, treating the porous media as lines of rectangular slits, with Sampson flows at low $Re$, and inertial corrections at finite $Re$.  

The present paper is structured as follows. In Section II, we present the experimental and numerical methods and compare the experimental measurements with those of matching simulations. In Section III, we report the results of systematic numerical simulations, focusing on the influence of the confinement on the flow rates and regimes. In Section IV, we rationalize the results at low and intermediate $Re$ assuming that the first line of cylinders controls the flow into the array. The model accounts for the viscous dissipation at the aperture with the Sampson permeability of rectangular slits at $Re \ll 1$, and inertial corrections starting at $Re \approx O(1)$. The results are consistent with the numerical and experimental data, as discussed in Section V. We conclude in Section VI with the implications of our findings and future work.



\section{EXPERIMENTAL AND NUMERICAL MODELS: methods, results, and comparison}\label{sec: Experimental methods}

In this section, we introduce the experimental system and present numerical simulations to characterize how the flow through a finite porous medium depends on its confinement. The model porous medium is a square array of cylinders in a closed channel. We investigate the dependence of the flow rate through the array on its porosity and confinement in creeping flows.

\subsection{Experimental system and procedure}\label{sec: Experimental methods}

The experimental set-up consists of a channel of rectangular cross-section, with an embedded porous medium, as represented in Fig. \ref{fig:expsetup}. The channel is assembled from acrylic sheets (3.17 mm thick), laser-cut to size with a precision of $\pm 0.1$ mm. The sides of the channel are glued with acrylic cement and sealed with clear waterproof silicone, ensuring that its dimensions are accurate $\pm 2$ mm and there is no leak. The channel is connected to a gear pump (Cole-Parmer Masterflex) and a flowmeter (McMillan Flow) in a closed-loop configuration, as shown in Fig.~\ref{fig:expsetup}. The velocity field is measured with Particle Image Velocimetry (PIV) in the mid-plane, $y = 0$. The flow is seeded with 20 $\mu$m polyamide particles ($\rho_p =1.03$ g/cm$^3$, Dantec Dynamics) at a volume fraction of 1 g/L, and the center plane of the channel is illuminated with a horizontal laser sheet (Dantec Dynamics RayPower 2000) approximately 1 mm thick. We track the particle displacements with a camera positioned directly above the cylinder array (Dantec Dynamics Flowsense USB). The images are recorded at frame rates between 10 and 60 fps, depending on the experiment. To vary the array confinement, we use 3 flow channels with widths $w = 40$~mm, 60~mm, and 120~mm. All the channels have a height $h=40$~mm, and a length $L \approx 2.1$~m (see Supplemental Material for a picture of an acrylic channel \cite{SM}). The pump sets the flow rate through the channel $q$, we measure the total flow rate $Q$, the flow rate entering the array $Q_{in}$, and the flow rate exiting the array downstream $Q_{out}$, in the midplane. From those measurements, we can infer the undisturbed flow rate $Q_0$ and the flow rate exiting the array, $Q_{side}$.

\begin{figure}
\centering\includegraphics[width=1\linewidth]{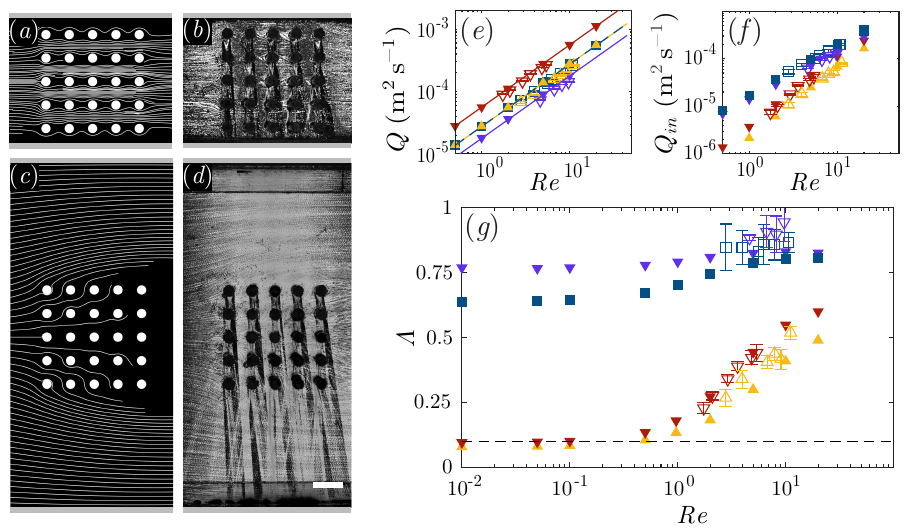}
\caption{Experimental characterization and numerical modeling of the flow through a confined array. \textmd{Streamlines at $Re = 6$ for a highly confined array, $C=0.88$, from (a) numerical simulations and (b) long exposure imaging of PIV particles. The dark and bright streaks are due to the interactions of the laser sheet with the cylinders. Streamlines at $Re = 5$ for a low confinement array, $C = 0.29$, from (c) numerical simulations and (d) long exposure imaging of PIV particles. In (a-d), the arrays have a porosity $\phi = 0.84$ and the scale bar correspond to 10 mm. Flow rates in the mid-plane, at $y = h/2$: (e) $Q$, flow rate across the channel width, (f) $Q_{in}$, flow rate into the array and (g) $\Lambda$, ratio of the flow rate into the array over the reference flow rate. The solid and hollow markers represent the numerical and experimental results, respectively, and the solid lines are the theoretical model. The dashed line corresponds to $\Lambda = 0.1$. The marker shape indicates the array porosity: $\square \; \phi = 0.89$, $\triangledown \; \phi = 0.84$, and $\vartriangle \; \phi = 0.73$. The color corresponds to the confinement {\color{dred}\rule[0.1em]{10pt}{4pt}} $C = 0.29$, {\color{yel}\rule[0.1em]{10pt}{4pt}} $C = 0.45$, {\color{dblue}\rule[0.1em]{10pt}{4pt}} $C = 0.72$, {\color{dpple}\rule[0.1em]{10pt}{4pt}} $C = 0.88$ \label{fig:expvsnum}}}
\end{figure}

The finite porous media are square arrays of $n\times n$ clear acrylic rods. The clear rods let the laser sheet through, yet the difference in index of refraction between acrylic and water lead to the formation of shadows, i.e. dark regions (see Fig. \ref{fig:expvsnum}). In those regions, the PIV measurements are not accurate or possible. We measure the velocity field around the half of the array closest to the laser and extrapolate the results using the symmetry of the system. We set the size of the array $n=5$, the cylinder diameter $d \approx 3.2$~mm (1/8 in.) and the height $h=40$~mm is equal to the channel height. We vary the center-to-center spacing of neighboring cylinders from $\delta = 6$ to 10 mm. The confinement of the array, $C$, is the ratio of the width of the array to the width of the channel:
\begin{equation}\label{eqn:Confinement}
 C= \cfrac{(n-1)\,\delta+d}{w} 
\end{equation}
The porosity $\phi$ is equal to the volume fraction of void space in the array. The array of cylinders is inscribed in a square of side length $L_s = (n-1) \, \delta + d$. The area of a cylinder is $A_c = \pi d^2/4$, and the porosity is equal to
\begin{equation}
 \phi= \frac{L_s^2 h - n^2 A_c h }{L_s^2 h} = 1 -\frac{n^2 A_c}{L_s^2}
\end{equation}
Upon substitution, we get:
\begin{equation}\label{eqn:Porosity}
 \phi=  1 -\frac{\pi}{4}\left(\frac{n d}{(n-1)\, \delta+d}\right)^2
\end{equation}
The porosity reaches a minimum when $\delta = d$, $\phi_{min} = 0. 21$. Although all values of $\phi \geq 0.21$ are accessible, we focus on higher porosity values, relevant to biological and engineered applications with $\phi \geq 0.5$.

Before each experiment, we insert the cylinder array into the channel base and seal it with silicone adhesive. The channel is then filled with water at room temperature, and air bubbles are removed from sealable ports on the top of the channel. During an experiment, we increase the flow rate from 0.1 to 1~L.min$^{-1}$, which corresponds to channel Reynolds numbers:
\[
Re_{channel} = \frac{\rho U \, D_h}{\mu} \in [21,416]
\]
with $U$ the average flow velocity, $D_h$ the hydraulic diameter of the channel $D_h = {2 w h}/(w + h)$, and $\rho$ and $\mu$ the density and dynamic viscosity of water, respectively. The cylinder array is positioned two-thirds of the channel length from the inlet, beyond the channel entry length, estimated as
$L_e  = 0.06 \times Re_{channel} \times D_h \leq 1 ~\text{m}$ \cite{Martinelli2019}. To further reduce entry and exit effects, porous diffusers or flow straighteners are at both ends of the channels: Each flow straightener consists of a vertical acrylic sheet with a hexagonal pattern of holes near the inlet and outlet of the channel, allowing the flow to pass through.

Once the flow rate is selected, we first check that the flow has reached the steady state by measuring the flow profile upstream of the cylinder array in horizontal planes, varying the vertical position of the laser in increments of 5 mm. We compare the measured profiles to COMSOL Multiphysics results for a steady Poiseuille flow in an infinite channel of identical cross section (see \ref{sec: Numerical Study} for a description of the numerical simulations and Supplemental Material for comparison of PIV measurements with the fully developed flow profile \cite{SM}). Once the velocity field in the horizontal planes has reached the steady state, we record PIV measurements in the mid-plane $y=0$ to reconstruct the flow streamlines, and evaluate the 2D flow rates through the channel and the array. The clear rods let the laser sheet through, yet the difference in index of refraction between acrylic and water leads to the formation of shadows, i.e., dark regions (see Fig. \ref{fig:expvsnum}). In those regions, the PIV measurements are not accurate or possible. We measure the velocity field around half of the array closest to the laser and extrapolate the results using the symmetry of the system. The 2D flow rate through the channel $Q$ is measured by integrating the velocity field over the line segment between $x_{-,c} = -w/2$ and $x_{+,c} = w/2$ at $y = 0$ at a distance $z$ equal to two array lengths, $2 \, L_s$ upstream from the array. The flow rate entering the array $Q_{in}$ is obtained upon integration of the velocity field over the line segment connecting $x_{-,a} = -\left((n-1)\,\delta +d\right)/2$ and $x_{+,a} = \left((n-1)\,\delta +d\right)/2$ at $y = 0$ at a distance $z$ equal to the radius of a cylinder, $d/2$ upstream from the array. Both flow rates are in m$^{2}$.s$^{-1}$.

\subsection{Numerical simulations}\label{sec: Numerical Study}
We use finite element analysis in COMSOL Multiphysics (Burlington, MA) to simulate flow through a confined array of cylinders. The numerical domain consists of a fully developed flow in an infinite channel of rectangular cross section ($h \times w$), with a $5 \times 5$ array of cylinders centered across the channel width and spanning the entire height. We solve the  continuity and Navier-Stokes equations for the Newtonian incompressible flow:
\begin{equation}\label{eqn:Continuity}
 \nabla \cdot \mathbf{u} = 0
\end{equation}
\begin{equation}\label{eqn:NavierStokes}
\rho \, \left[\frac{\partial \mathbf{u} }{\partial t} + \left(\mathbf{u}\cdot \nabla\right) \mathbf{u} \right]= \mu \nabla^2\mathbf{u}-\nabla p 
\end{equation}
where $\mathbf{u}$ is the vector velocity and $p$ the pressure. The density $\rho$ and viscosity $\mu$ are equal to those of water, i.e. $\rho = 10^3$ kg.m$^{-3}$ and $\mu = 10^{-3}$ Pa.s.
We impose the following boundary conditions:
\begin{eqnarray}\label{eqn:BC}
     \mathbf{u} &=& 0 \hspace{0.5cm} \mbox{on cylinder walls}\\
     \mathbf{u} &=& 0 \hspace{0.5cm} \mbox{on domain walls} \; (\pm\frac{1}{2}w \; \mbox{and} \; \pm\frac{1}{2}h)
\end{eqnarray}
 In all simulations, the channel and cylinder height are $h = 40$ mm, and the cylinder diameter is $ d = 3.2$ mm. The other geometrical properties of the system are varied systematically with the channel width $w$ ranging from 26 to 120 mm and the cylinder-to-cylinder spacing $\delta$ from 5 to 10 mm. The simulations use a fully developed inlet condition with an average velocity $U = q/(hw)$, where $q$ and $w$ are set to the experimental values for direct comparison. Additional simulations are conducted to explore a wider range of flow properties, including the pressure, which is not measured experimentally, as we systematically vary $U$, $\delta$, and $w$. All simulations are at steady state and in three dimensions. To allow direct comparison with the PIV measurements, all numerical data are reported at the mid-plane of the channel, $y = 0$. The Reynolds number of the cylinders:
 \begin{equation}\label{eqn:Re}
 Re = \frac{\rho d U}{\mu} 
\end{equation}
varies between $10^{-3}$  and $45$ to study low and intermediate Reynolds number flows. The inlet velocity is therefore defined as
  $U  = \mu Re/\left(\rho d \right) \in [3.15 \times 10^{-7}, 1.47 \times 10^{-2}]$ m.s$^{-1}$. \\
  We mesh the channel to solve the conservation equations while imposing the boundary conditions. The physics-controlled mesh elements consist of tetrahedra, prisms, triangles, and quadrilaterals, whose density adapts to the flow and is smallest near boundaries.  The total number of elements ranges from about $15\times 10^6$ to $20\times 10^6$, depending on the channel geometry. To ensure accurate simulations, a mesh-refinement study is run for a 60 mm-wide channel with $\delta = 6$ mm. This study shows that simulations with finer ($15.1 \times 10^6$ elements) and coarser mesh ($4.4 \times 10^6$ elements) produce flow rate values within 0.5\% of the chosen mesh with $15.09 \times 10^6$ elements. We also tested the sensitivity of the simulations to the experimental uncertainty on the dimensions of the channel. At Reynolds number $Re=40$, the relative error in $Q_{in}$ is below  $1\%$.


\subsection{Experimental and numerical results and comparison}\label{sec:Experimental and numerical results}
 Both experiments and numerical simulations show that the flow rate entering the array increases with the confinement. This can be observed by comparing the flow into arrays of identical porosity placed in channels of different widths. In Figs. \ref{fig:expvsnum}(a-d), we report the streamlines produced by overlapping PIV images and from numerical simulations. We compare the flow through an array of porosity $\phi =0.84$ under high confinement C = 0.88 in Figs. \ref{fig:expvsnum}(a-b) and under low confinement C = 0.29 in Figs. \ref{fig:expvsnum}(c-d).  At high confinement, the streamlines flow straight through the array with little curvature. This is consistent with the sieve regime. At low confinement, the particle path curves around the array. The particles that enter the array quickly exit the sides of the array. This array is in the deflection regime. Despite having similar Reynolds numbers and identical porosity, the arrays exhibit different flow regimes. These qualitative observations provide experimental and numerical evidence of the influence of the confinement, which is now studied more quantitatively through flow rate measurements.  

The flow characterization experiments include two sets of conditions. First, the spacing $\delta$ is kept constant ($\delta = 8$ mm) while the width of the channel varies from $w = 40$ to $w = 120$ mm. Second, The porosity of the array is equal to $\phi = 0.84$, and the confinement varies between $C = 0.29$ and $C = 0.88$.  For both systems, the flow rate measured across the channel in the mid-plane $Q$ increases linearly with the Reynolds number of the flow $Re$ and the width of the channel $w$ (see Fig. \ref{fig:expvsnum}(e)). The experimental results are in good agreement with the numerical simulations, and with the flow rate obtained by integrating the velocity field of a Poiseuille flow in a rectangular channel (see \cite{Papa2021, SM} for derivation of $Q$). Whereas the flow rate in the channel is larger for the wider channel, i.e. in the low confinement geometry, the flow rate entering the array $Q_{in}$ is larger for the narrower channel, i.e. in the high confinement geometry (see Fig. \ref{fig:expvsnum}(f)). This measurement is consistent with the observations presented in Figs. \ref{fig:expvsnum}(a-d): the confinement increases the amount of fluid entering the array. 

In previous work, a dimensionless flow rate entering the array called leakiness $\Lambda$ is defined by the ratio of $Q_{in}$ and $Q_0$, where $Q_0$ is the flow rate through the same region of the channel with no cylinder \cite{Cheer1987, Barta2006, Koehl2013}:
\begin{equation}
\Lambda = \frac{Q_{in}}{Q_0}
\end{equation}
Although in our system one could use the flow rate across the channel width to non-dimensionalize the entering flow rate, this definition would not be practical for comparing our results with the previous work on unbound systems. The flow rate through a centered region of width $(n-1) \delta +d$ or $4 \delta +d$ is related to the flow rate across the channel. The ratio between $Q_0$ and $Q$ is obtained analytically from the velocity of the Poiseuille flow (see derivation in the Supplemental Material \cite{SM}). With $Q_0 = \alpha' Q$, we can define $\Lambda$ from the measurement of $Q$ and $Q_{in}$:
\begin{equation}
\Lambda = \frac{Q_{in}}{\alpha'Q}
\end{equation}
The ratio $\Lambda$ is plotted on Fig. \ref{fig:expvsnum}(g). The experiments show that the leakiness of the array of porosity $\phi = 0.84$ increases with both $Re$ and confinement $C$. The leakiness increases by a factor two from 0.25 to 0.5, over the range of $Re$ explored ($Re\in [3,11]$), for the low confinement system.  Yet the leakiness of the low-confinement system remains lower than that of the high-confinement geometry, which is $\approx 0.7$. These results are consistent with numerical simulations, which allow for exploring a larger range of $Re\in[5\times 10^{-2}, 20]$. All systems experience their lowest leakiness at low $Re$. In the low-confinement system, the leakiness is below the threshold of $\Lambda = 0.1$, which corresponds to the rake regime. At high confinement, the leakiness is systematically above the threshold. Both systems exhibit an increase in leakiness with increasing $Re$, reaching a plateau at $Re \approx 10$. The dependence of the leakiness on $Re$ is more pronounced at low than at high confinement. These results show the influence of confinement on the flow through and around the array and demonstrate an increase in leakiness with greater confinement. 

In practice, both porosity and confinement are commonly found to vary simultaneously. Indeed, when the channel geometry is unchanged but the porosity or spacing between the cylinders varies, both porosity and confinement change \cite{Hood2019}. In a second set of experiments, we consider this practical situation,  with the width of the channel $w = 60$ mm while the spacing $\delta$ increases from 6 to 10 mm. In addition to varying the porosity from 0.73 to 0.89, these geometries yield different confinement values, $C=0.45$ and $C= 0.72$, respectively. As both porosity and confinement increase, the differences in leakiness are more pronounced. The low porosity and low confinement system experiences very low leakiness at low $Re$. The leakiness then increases with $Re$ and reaches intermediate values (about 0.5) at $Re \approx 40$. The high porosity and high confinement system exhibits large leakiness at low $Re$, with values exceeding 0.5. The leakiness then increases marginally ($\approx 10\%$) with $Re$. These results illustrate scenarios in which the porosity of the array enhances the effect of confinement.

The experimental results demonstrate the influence of the Reynolds number $Re$, porosity $\phi$, and confinement $C$ on the transport through the array. The comparison with numerical results shows good agreement with the experimental data, with an average relative error of 12.3 $\%$, and reveals the system's response at low $Re$ values, which are not experimentally tested due to pump limitations, as the flow becomes unsteady at low flow rates. We therefore rely on numerical simulations to study the effect of confinement on the leakiness and flow regimes. We can systematically vary the system geometry (through $\delta$ and $w$) and determine the velocity field and the pressure, which would be difficult to measure experimentally. 



\begin{figure}
\centering\includegraphics[width=0.95\linewidth] {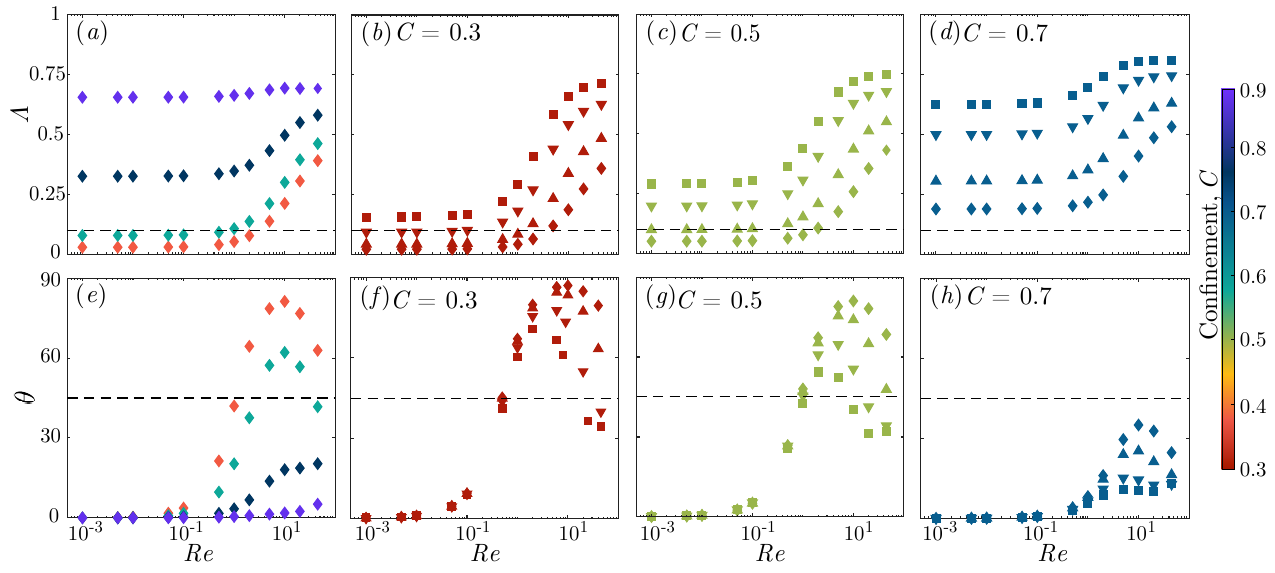} 
\caption{\textmd{Influence of confinement on flow through array characterized by $\Lambda$ in ($a-d$) and the flow angle $\theta$ in ($e-h$). In ($a$) and ($e$), the porosity is equal to $\phi = 0.63$ and confinement $C = [0.38, 0.58, 0.77, 0.89]$.  The confinement is equal to $C = 0.3$ in ($b$) and ($f$), $C = 0.5$ in ($c$) and ($g$), and $C = 0.7$ in ($d$) and ($h$). The porosity is indicated by the marker shapes: $\square \; \phi = 0.89$, $\triangledown \; \phi = 0.84$, $\vartriangle \; \phi = 0.73$, and $\Diamond \; \phi = 0.63$. The dashed lines indicate threshold values $\Lambda = 0.1$ for ($a-d$) and $\theta = 45^o$ for ($e-h$). }}
 \label{fig:num}
\end{figure}
 

\section{Influence of confinement on flow through arrays of cylinders}\label{sec: FLOW REGIMES THROUGH FINITE POROUS MEDIA}

We conduct two types of numerical simulations. In the first set of simulations, the porosity of the array is fixed to an intermediate value $\phi=0.63$, and the confinement varies, $C \in [0.38,0.89]$. The second simulations consider systems of fixed confinement $C=0.3$, $C=0.5$, and $C=0.7$ while varying the porosity from $\phi=0.63$ to $\phi = 0.89$. As in Section \ref{sec:Experimental and numerical results}, we report the leakiness $\Lambda$ (see Figs. \ref{fig:num}(a-d)).

\subsection{Leakiness at low $Re$: definition of the rake regime}\label{subsec: leakiness}
For all geometries, we report that leakiness increases with $Re$. The leakiness is minimum and constant for $Re\ll 1$ in the Stokes regime. At low confinement and low porosity values, the leakiness at low $Re$ is smaller than $\Lambda = 0.1$. This means that over $90\%$ of the fluid originally traveling through the region occupied by the array is diverted into the side channels, around the array. This is the rake regime. As confinement and porosity increase, the minimum leakiness can reach 0.6, so that the majority of the fluid is not diverted around the array. This result indicates that systems with high porosity and confinement do not experience the rake regime. This can be advantageous for filtration processes, but would be associated with lower drag on bristled wings through which the flow should be limited. The range of porosity and $Re$ in which a system experiences the rake regime depends on the confinement. As the confinement increases, the range of porosity experiencing the rake regime decreases: The porosity at which the system leaves the rake regime decreases as confinement increases. For $C = 0.3$, porosity values up to 0.84 are in the rake regime. For $C = 0.5$, porosity values lower than 0.77 exhibit the rake regime. At high confinement, no porosity in the 0.63 to 0.89 range experiences the rake regime. For a given porosity, for example $\phi = 0.63$, the range of $Re$ for which the rake regime is observed decreases as the confinement increases. In Fig. \ref{fig:num}(a), we observe that the leakiness of the system becomes larger than 0.1 at $Re = 2$ for $C = 0.38$ and at $Re = 0.7$ for $C = 0.58$. As the confinement increases, the width of the side channel decreases, forcing more fluid into the array. For a given confinement, systems with increasing porosity offer less resistance to flow through the array, thereby increasing leakiness. 

At intermediate $Re$, $Re \geq 0.1$, the leakiness increases rapidly as a function of $Re$, except for the high confinement and high porosity systems, for which the leakiness at low $Re$ is already large, $\Lambda \geq 0.6$, and increases only marginally. As $Re$ increases, if leakiness increases to 0.7, it becomes constant for $Re \geq 10$ as observed in Figs. \ref{fig:num}(a-d). Increasing $Re$ increases the fraction of fluid that travels through the array. This is consistent with previous work that rationalized this observation by noting the decrease of the boundary layer thickness on the cylinders as $Re$ increases \cite{Hood2019}. Yet, the relative increase in leakiness $\Lambda(Re=45)/\Lambda(Re \ll 1)$ depends on porosity and, more significantly, on confinement. At high confinement, for all porosity, the increase is limited. For example at $C = 0.89$ and $\phi =0.63$, $\Lambda(Re=45)/\Lambda(Re \ll 1) = 1.06$. Indeed, at large confinement, a large fraction of the flow is forced into the array by the narrow side channels even at low $Re$. For an intermediate confinement $C= 0.5$, the relative increase in leakiness varies between 2.5 and 8.5 as the porosity decreases from 0.89 to 0.63.  At large porosity, the leakiness at $Re \ll 1$ is already large because the cylinders are far apart; the relative increase in leakiness is therefore smaller. Finally, for low confinement values, such as $C = 0.3$, the increase in leakiness reaches 17.5 for a porosity of $\phi =0.63$. This result is in agreement with the work done on non-confined cylinders \cite{Cheer1987}, which reported a large increase in leakiness. For all systems considered, the maximum leakiness ranges from 0.35 to 0.75. As a result, all configurations are in either the deflection or sieve regime at $Re = 45$. 

In summary, for $Re$ between $10^{-3}$ and 45, the leakiness is an increasing function of porosity, confinement, and $Re$. The low-porosity and low-confinement arrays experience the rake regime at $Re\leq1$. The high-porosity and high-confinement arrays do not experience the rake regime, as the relative amount of fluid entering the array is non-negligible.  All arrays experience the sieve or deflection regime at $Re>10$. The relative change in leakiness over the range of $Re$ varies between $1.1$ and $17$, with the low-confinement, low-porosity arrays experiencing the largest relative increase in leakiness.

\subsection{Flow angle: deflection and sieve regimes}\label{subsec: flowangle}
For finite values of $Re$, we characterize the flow inside the array. Figures \ref{fig:num}($e-h$) report the angle of the flow $\theta$ defined by its tangent: 
\begin{equation}
\tan \theta = \frac{Q_{side}}{Q_{out}} = \frac{Q_{in}- Q_{out}}{Q_{out}} 
\end{equation}
where $Q_{side}$ is the flow rate exiting the array through the sides in the x-direction and $Q_{out}$ is the flow rate exiting the array, downstream, in the z-direction (see Fig. \ref{fig:expsetup}). Using mass conservation, we know that $Q_{side} = Q_{in}-Q_{out}$. The angle of the flow is the angle between the average velocity in the array and the $ z$-direction in the upper half of the array ($x>0$). By symmetry, the angle is equal and opposite for the lower half of the array ($x<0$). Values of $\theta$ vary between $\theta = 0^o$ when there is no side flow, and $\theta = 90^o$ when all the flow is diverted sideways, and no fluid exits the array in the $z-$direction. The intermediate value of $\theta = 45^o$ corresponds to half of the fluid leaving the array through its side and the other half downstream: $Q_{side} = Q_{out} = Q_{in}/2$. For angles $\theta < 45^o$, the amount of fluid exiting the array downstream is larger than the amount of flow exiting the array through its sides. For angles $\theta > 45^o$, the majority of the fluid leaves the array through the sides, before reaching the last line of cylinders. The dependence of the flow angle on $Re$ is non-monotonic. In the Stokes regime, $Re \leq 0.1$, the angle is equal to zero. Regardless of the value of the leakiness, $Q_{in} \approx Q_{out}$, at low $Re$. As a result, arrays can only be in the rake or sieve regime, depending on the leakiness. 

We now consider $Re>0.1$. For all geometries, the flow angle increases as $Re>0.1$, so $Q_{out}<Q_{in}$ when inertial effects are no longer negligible. For the high porosity and high confinement geometries, typically with $C>0.7$ and $\phi \geq 0.84$, the flow angle increases and reaches a maximum at a small value $\theta \lesssim 20^o$. This is consistent with the large leakiness of those systems: the resistance to the flow inside the array is low compared to the resistance to the flow around the array, and so the fluid travels across the array. For lower confinements, arrays of all porosity experience an increase in flow angle until $Re \approx 10$. The maximum angle can reach $90^o$ for low confinement and low porosity. In this case, the resistance to flow is high inside the array and low outside, leading to curved streamlines and significant side flow.  The maximum flow angle decreases as the confinement and porosity increase. If the maximum value is lower than $45^o$, the array can only be in the rake and sieve regimes. If the maximum value is larger than $45^o$, the array experiences the deflection regime. As $Re$ increases, above $Re = 10$, the flow angle decreases. For arrays with a maximum flow angle below $45^o$, the decrease in flow angle is not associated with a regime change. For arrays with a maximum flow rate larger than $45^o$, the flow angle can remain above $45^o$, typically for low confinement and low porosity (see for example $C = 0.3$ and $\phi = 0.63$) or become lower than $45^o$, as observed for intermediate confinement and porosity (see for example $C = 0.5$ and $\phi = 0.84$). For such values of $Re$, inertia drives the fluid across the array, reducing the flow angle to values that reflect the degree of flow hindrance in the array. This decrease in angle can result in a transition from the deflection to the sieve regime for the systems that are most favorable to flow in the array, with higher porosity and confinement.

\begin{figure}
\centering\includegraphics[width=1\linewidth]{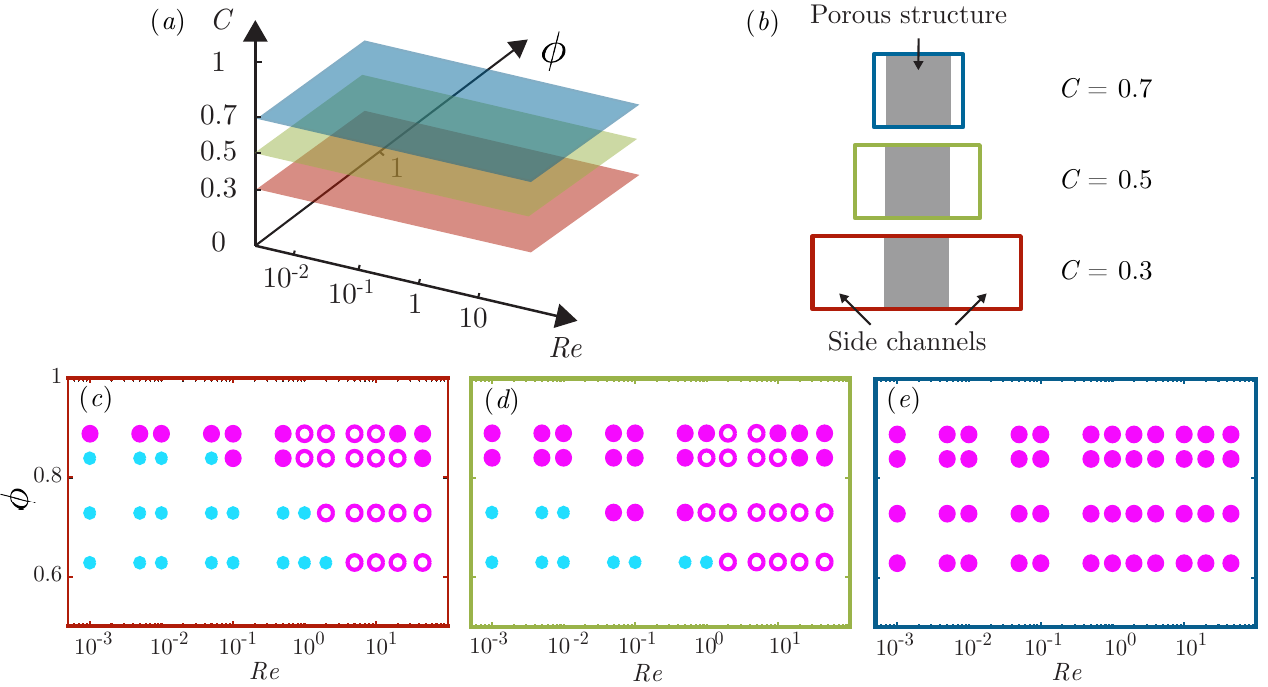}
\caption{\textmd{Regime diagram. (a) The flow regime through the array depends on porosity $\phi$, Reynolds number $Re$, and confinement $C$. (b) Schematic of the channel cross-section with the area occupied by the porous structure in gray for the three confinement values $C = [0.3, 0.5, 0.7]$}. The regimes diagrams for $C=0.3$ (c), $C = 0.5$ (d), and $C = 0.7$ (e)  exhibit three flow regimes: {\color{lblue}$\bullet$} rake, {\color{lple}$\medcirc$} intermediate, and {\color{lple}$\medbullet$} sieve. }\label{fig:RegimeDiagram}
\end{figure}
 \definecolor{lblue}{HTML}{21deff} 
\definecolor{lple}{HTML}{f50aff} 

\subsection{Regime diagrams}\label{subsec: phasediagram}

We summarize our findings for the low, intermediate, and high confinement systems ($C =$ 0.3, 0.5, and 0.7, respectively) in regime diagrams, shown in Fig. \ref{fig:RegimeDiagram}. At low $Re$, $Re \lesssim 0.1$, the cylinder arrays are in the rake or sieve regimes. The range of porosity associated with the rake regime decreases as the confinement increases, eventually disappearing. At high confinement, all porosity values are in the sieve regime. At low confinement, a wide range of porosities falls within the rake regime. As porosity increases, the array exits the rake regime at lower $Re$. 

At intermediate $Re$, $0.1 \lesssim  Re \lesssim 10$, we focus on the low and intermediate confinements. The arrays with lower porosity transition from the rake to the deflection regime and remain in that regime. The intermediate porosity arrays experience the sieve regime at $Re \leq 1$ before the deflection regime. In this case, the system has a low leakiness at low $Re$. At intermediate $Re$, both leakiness and flow angle increase above their threshold values, $0.1$ and $45^o$ respectively, resulting in this combination of sieve and deflection regimes. At high porosity, the system transitions from the low-$Re$ sieve regime to the deflection regime as the flow angle increases above $45^o$. 

Eventually, for $Re>10$, the systems remain in the deflection regime or, if the porosity is sufficiently large, return to the sieve regime at $Re \approx 40$.

In summary, under high confinement, an array of cylinders only experiences the sieve flow regime, which means that the relative flow rate through the array is non-negligible ($\Lambda >0.1$) and most of the flow entering the array travels across the length of the array. At intermediate and low confinements, the arrays can experience all three flow regimes previously reported: rake at low $Re$, deflection at intermediate $Re$, and sieve at low and larger $Re$, depending on the porosity of the array. To rationalize the complex interplay between confinement, porosity, and $Re$, in a system with a square array of cylinders of diameter $d$ and height $h$, we propose a theoretical description which models the flow in the pores, or channels bound by two neighboring cylinders and in the side channels, or channels between an outer cylinder and the wall of the channel.

\section{Permeability-based model of the flow}\label{sec: Finite porous medium model}
We rationalize our results by comparing them to an analytical solution for Darcy flows within and around the array. The permeability model is based on aperture flow through slits with inertial corrections \cite{Jensen2014}.

\begin{figure}
\centering\includegraphics[width=0.9\linewidth]{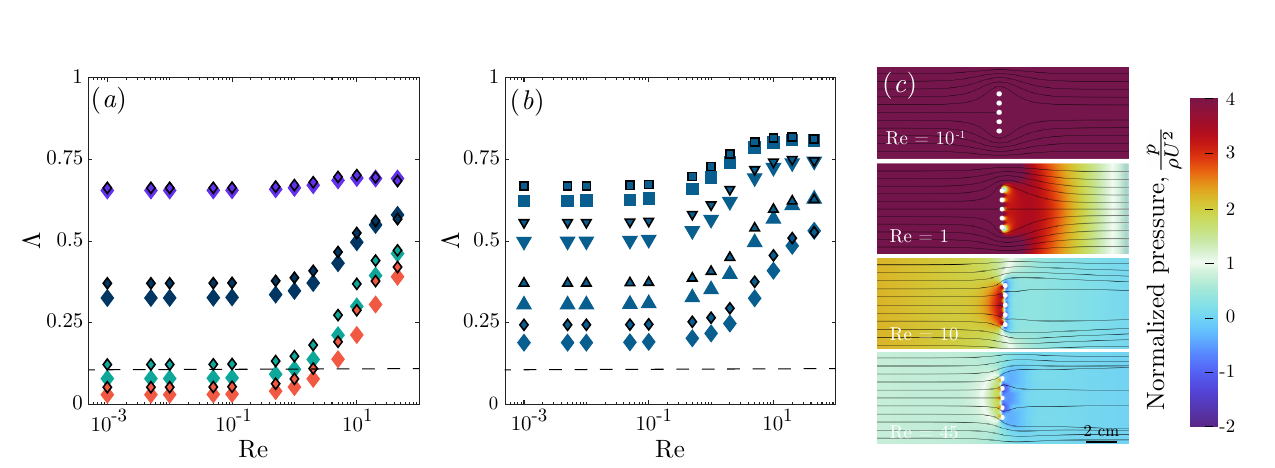}
\caption{\textmd{Flow through a single column of cylinders. In ($a$), the porosity is $\phi = 0.63$ and the confinement $C = [0.38, 0.58, 0.77, 0.89]$, from bottom to top, using the same color scheme as in Fig \ref{fig:num}.  In ($b$), the confinement is equal to $C = 0.7$ and the porosity is indicated by the marker shapes: $\square \; \phi = 0.89$, $\triangledown \; \phi = 0.84$, $\vartriangle \; \phi = 0.73$, and $\Diamond \; \phi = 0.63$. The dashed lines indicate threshold values $\Lambda = 0.1$ for ($a-b$). The large markers correspond to the $5 \times 5$ array, the small markers to the column or $5 \times 1$ array. (c) Rescaled pressure field for a column of 5 cylinders with $\phi = 0.73$ and $C = 0.45$. The rescaled pressure is equal to $P/\rho U^2$. The black lines are representative streamlines. \label{fig:singlerow}}}
\end{figure}

The leakiness of the array, $\Lambda$, is an increasing function of the Reynolds number, which depends on confinement and porosity. The increase in leakiness with $Re$ was first predicted in an unconfined geometry composed of two infinite cylinders by Cheer and Khoel \cite{Cheer1987}. This result indicates that the leakiness of a pair of cylinders exhibits a similar behavior to that of an array of 25 cylinders. To further this comparison, we conduct numerical simulations with arrays composed of a single column of 5 cylinders, i.e. 5 cylinders in a line perpendicular to the direction of the flow. We model arrays with a fixed porosity $\phi= 0.63$ and confinement values from $C = 0.38$ to $C= 0.89$, and a fixed confinement $C = 0.7$ with varying porosity $\phi \in [0.63,\, 0.89]$, see Figs. \ref{fig:singlerow}(a-b). Those single-column systems exhibit an increase in leakiness with increasing $Re$, as do the $5 \times 5$ arrays. In addition, systems composed of a single column of five cylinders and a $5 \times 5$ square array of cylinders, with the same porosity and confinement, exhibit very similar leakiness. One can therefore conclude that the leakiness of the array is set mostly by its first column or line of cylinders.

The geometry of a single line of cylinders is analogous to the structure of a thin plate with holes or a porous membrane. Filtration membranes have motivated numerous studies on fluid transport through pores and drag force on porous plates \cite{Gravelle2013, Jensen2014, Strong2019, Hadjaje2025}. Both early and recent studies have focused on the flow through a single pore. The seminal work by Sampson \cite{Sampson1891} defines the pressure drop associated with the flow near the entrance of the pore and derives an analytical solution for the permeability of a circular single pore in an infinite membrane of zero thickness at low $Re$. Roscoe then derived the permeability of slit rectangular-shaped pores similar in shape to the opening between neighboring cylinders \cite{Roscoe2009}. For pores of finite thickness, an additional pressure drop due to the viscous dissipation in the pore can be added. The linear superposition of the hydraulic resistances of the aperture (Sampson resistance) and the channel flow (Poiseuille resistance) has been validated for different pore geometries \cite{Dagan1982, Jensen2014}. Recent work investigated the inertial correction to the Sampson model for moderate $Re \in [10^{-1}- 200]$. Following Bernoulli's principle, the inertial contribution to the pressure drop across the pore increases as $\rho u_p^2$, where $u_p$ is the velocity in the pore \cite{Jensen2014, Hadjaje2025}. It is worth noting that hydrodynamic interactions between neighboring pores tend to increase the permeability of an isolated pore. In this study, the pores or slits will be assumed thin and independent.

We first model the leakiness of the line of cylinders at low $Re$, for $Re \leq 10^{-1}$, a regime in which leakiness is constant and only depends on the confinement and porosity. The leakiness at low $Re$ is minimum, indicating whether the system can reach the rake regime, if $\Lambda_{min} \leq 0.1$. We then present a model that captures the effects of inertia with an additional pressure drop across the pore and the formation of a low-pressure wake behind the cylinders.



\subsection{Stokes flow and Sampson approximation}
 At low Reynolds numbers, we neglect inertial effects and consider the incompressible Stokes flow through and around the array. The ($n-1$) pores between the cylinders and the 2 side channels between the array and the side walls are approximated with rectangular slits of height $h$ and widths that depend on the confinement and porosity. Roscoe defined the linear dependence of the pressure drop $dp$ on the flow rate $q$ through an elliptical and a rectangular slit of width $a$, height $b$, with $a << b$, and zero thickness \cite{Roscoe2009}:

 \begin{equation}\label{eqn:Darcy}
dp = R_{slit} q 
 \end{equation}
with 
 \begin{equation}\label{eqn:Roscoe}
R_{slit}  = \frac{32 \mu}{\pi a^2\,b} 
 \end{equation}
where $R_{slit}$ is the hydraulic resistance of the slit permeability. The hydraulic resistance in Eq. \ref{eqn:Roscoe} is derived for a single slit in an infinite plate. In first approximation, we assume that the distribution of the velocity in the channel does not significantly modify the resistance of a rectangular slit, which can be at the center of the channel or near the wall. We define the hydraulic resistance of the two types of rectangular slits in the system, i.e. the pores between cylinders, noted $P$, and side channels, noted $SC$:
\begin{eqnarray}
  \label{eqn:slitsthinP}  R_{P}&=&\frac{32 \mu}{\pi h(\delta-d)^2} \\
\label{eqn:slitsthinSC} R_{SC}&=& \frac{32 \mu}{\pi h w_{s}^{2}} 
\end{eqnarray}
where $2 w_s = w-(n-1)\,\delta-d$. These relations assume that $\{w_s, \, \delta-d\} < h$, which is the case for most confinement and porosity values relevant to biological and engineered processes and the systems considered here. The zero-thickness approximation is based on the magnitude of the viscous dissipation in the Poiseuille flow through the pores and side channels. For geometries considered here, the dissipation associated with the Poiseuille flow is mostly negligible compared to that of the Sampson flow.

We assume that each slit is independent and define the  hydraulic resistance of the line of $n$ cylinders, $R$:
\begin{equation}\label{eqn:R}
    R=\frac{R_{{P }}\, R_{{SC }}}{(n-1) R_{{SC }}+2 R_{{P }}}
\end{equation} 

The pressure drop across the column of cylinders $dp$ is
\begin{equation}\label{eqn:Darcy}
dp=R \, q
\end{equation}
where $q = (n-1)q_P + 2 q_S = h w \, U$ is the flow rate through the cross section of the channel. To define the leakiness of the cylinder line, we define the flow rate through the pores:
\begin{equation}\label{eqn:qin}
q_{in}  = (n-1) q_{{p}}=(n-1) \frac{dp}{R_{p}} = (n-1) \frac{R}{R_{P}} q.
\end{equation}

In summary, the relative flow rate into the array depends on the relative hydraulic resistance or permeability of the pores and the side channels:
\begin{equation}
\frac{q_{in}}{q}=(n-1)\frac{R}{R_{P}}.    
\end{equation}
Using the geometrical parameters of the system, we get:
\begin{equation}\label{eqn:qinq}
\frac{q_{in}}{q}=\left(1+\frac{2}{n-1} \frac{w_{s}^{2}}{\left(\delta-d\right)^{2}}\right)^{-1},
\end{equation}
with 
\begin{eqnarray}
\label{eqn:ws} w_s &=& \frac{1-C}{C}\frac{nd}{4} \sqrt{\frac{\pi}{1-\phi}},\\
\label{eqn:delta} \delta &=& \frac{d}{n-1}\left[ \frac{n}{2} \sqrt{\frac{\pi}{1-\phi}}-1\right].    
\end{eqnarray}
We can then express the relative flow rate into the array as a function of the dimensionless parameters of interest, the array porosity $\phi$, and confinement $C$, using Eqs. \ref{eqn:Confinement} and \ref{eqn:Porosity}: 
\begin{equation}\label{eqn:LowReModel}
    \frac{q_{in}}{q} = \left[1+\frac{(n-1)\,\pi\,(1-C)^2}{2C^2 \, \left(\sqrt{\pi} -2\sqrt{1-\phi}\right)^2} \right]^{-1}.
\end{equation}
We thus conclude that  leakiness of the array is defined as 
\begin{equation}\label{eqn:LambdaLowReHighC}
    \Lambda = \frac{q_{in}}{q_0} = \frac{q_{in}}{\alpha q} 
    = \frac{1}{\alpha}\left[1+\frac{(n-1)\,\pi\,(1-C)^2}{2C^2 \, \left(\sqrt{\pi} -2\sqrt{1-\phi}\right)^2} \right]^{-1},
\end{equation}
with the coefficient $\alpha$ defined analytically for a Poiseuille flow in a rectangular channel (see Supplemental Material for the derivation and expression \cite{SM}). This equation is derived based on the assumption that all slits have widths $\{w_s, \delta - d\} \leq h$. At low confinement, e.g. $C = 0.1$, the side channels can have a width $w_s$ larger than the height. If $w_s>h$ and $\delta -d\leq h$, Eqs. \ref{eqn:slitsthinP}, \ref{eqn:slitsthinSC}, and \ref{eqn:LambdaLowReHighC} become:
\begin{center}
\begin{eqnarray}
  \label{eqn:RPLowReLowC}     R_{P}&=&\frac{32 \mu}{\pi h(\delta-d)^2}, \\
\label{eqn:RSCLowReLowC} R_{SC}&=& \frac{32 \mu}{\pi h^2 w_{s}},\\
\label{eqn:LambdaLowReLowC}     \Lambda = \frac{q_{in}}{q_0}  = \frac{q_{in}}{\alpha q} 
   &=& \frac{1}{\alpha}\left[1+\frac{(n-1)}{n}\,\frac{2h}{d}\,\frac{1-C}{C} \,\frac{\sqrt{\pi(1-\phi)}}{\left(\sqrt{\pi}-2\sqrt{1-\phi}\right)^2}\right]^{-1},
\end{eqnarray}\label{LowRelowC}
\end{center}
respectively. We note that, at low porosity and low confinement, both the widths of the side channel and pore can be larger than $h$. This situation can be captured with the same approach, yet it is not relevant to the systems considered experimentally and numerically and will not be discussed here.

\begin{figure}
\centering\includegraphics[width=0.9\linewidth]{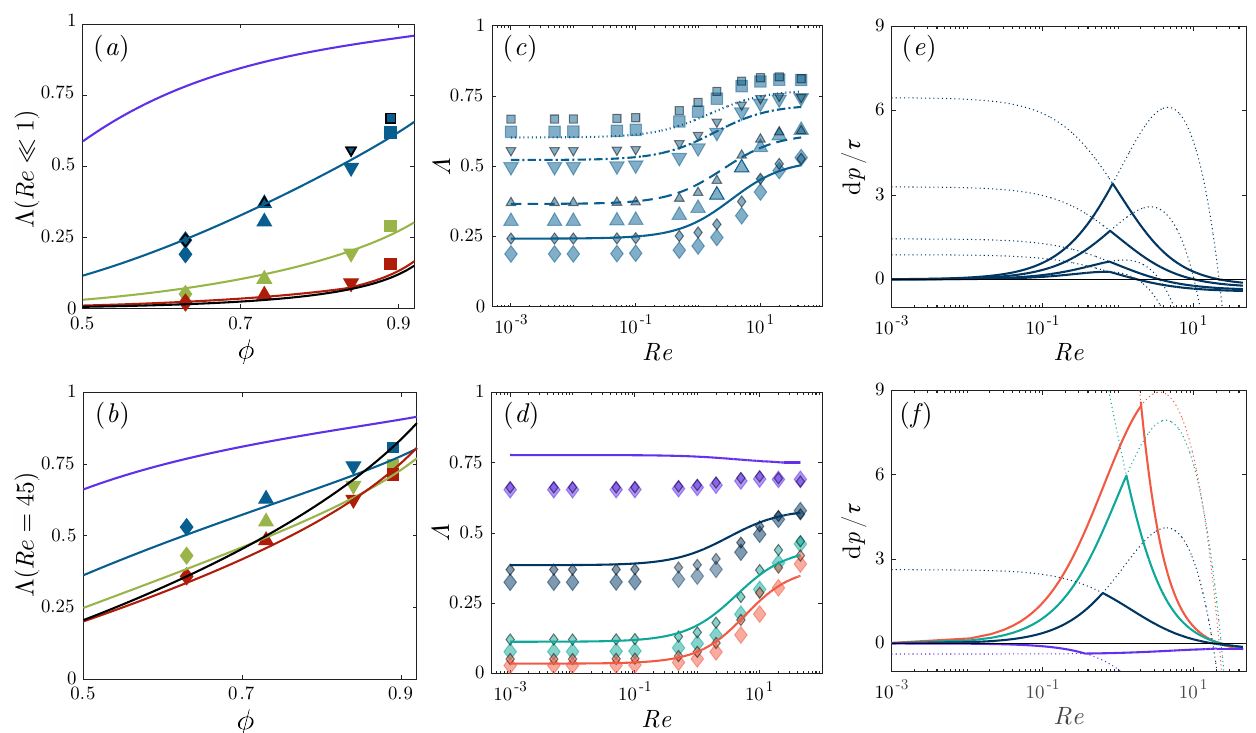}
\caption{\textmd{Leakiness at (a) low $Re$  and (b) $Re = 45$, for confinement values $C \in [0.1,\, 0.3,\, 0.5,\, 0.7,\, 0.9]$ using the same color scheme as in Fig. \ref{fig:num} and the black curve corresponds to $C=0.1$. The solid lines represent the model, the large markers the numerical results for the $5\times 5$ array and the small marker the numerical results for the $5\times 1$ array. Leakiness as a function of $Re$ for columns of cylinders with (c) a confinement $C = 0.7$ and $\phi= [0.63, 0.73, 0.84, 0.89]$ and (d) a porosity is $\phi = 0.63$ and $C = [0.38, 0.58, 0.77, 0.89]$. The solid lines correspond to the model. The markers are the same numerical results as in Fig. \ref{fig:singlerow}. In (e) and (f), the estimated inertial pressure difference between the pore and the side channel is rescaled with the viscous stress in the pore, until it becomes equal to the inertial pressure in the pore, and then it is rescaled with the inertial pressure in the pore (solid line). In (e), the columns of cylinders have a confinement $C = 0.7$ and $\phi= [0.63, 0.73, 0.84, 0.89]$. In (f) the porosity is $\phi = 0.63$ and $C = [0.38, 0.58, 0.77, 0.89]$.}}\label{fig:model}
\end{figure}

Despite simplifications to the pore geometry, the Sampson model captures the dependence of leakiness on porosity and confinement at low Reynolds number. To compare the results of the Sampson flow model with the numerical values, we plot the leakiness at $Re=0.01$ as a function of the porosity for different confinements. Fig. \ref{fig:model}(a) shows that the leakiness increases with both porosity and confinement. The quantitative agreement of the model with the numerical simulations indicates that the pores and side channels can indeed be modeled as thin rectangular slits in first approximation. The geometry of the cylinders and the viscous dissipation in the pores could be considered to improve the model accuracy. Yet the current model allows to predict the rake regime, which is defined for a leakiness lower than $10 \, \%$. As the confinement increases, the range of porosities yielding low leakiness is reduced. At low confinement, for example  $C = 0.1$ and $C= 0.3$, porosities up to $\phi \approx 0.9$ are in the rake regime, which is consistent with results obtained in unconfined \cite{Koehl2013} or low confinement geometries \cite{Hood2019}. At high confinement, $C \geq 0.7$, all arrays, regardless of their porosity, exhibit large leakiness at low Reynolds numbers, with $\Lambda > 0.1$, so they do not enter the rake regime. 
For low-Reynolds-number flows, the Sampson approximation with independent rectangular pores captures the leakiness of a line and an array of cylinders. At intermediate Reynolds numbers, the leakiness becomes an increasing function of $Re$, indicating the influence of inertia on the transport through and around the cylinder array.


\subsection{Inertial corrections}
At intermediate Reynolds numbers, inertial effects are no longer negligible. Inertia effectively reduces the permeability of an orifice, or increases the hydraulic resistance \cite{Hadjaje2025}. The inertial correction to the Sampson law is obtained from energy conservation. The pressure drop near the entrance is set by the sum of viscous and inertial losses.  To model the  intermediate Re, which are up to a few hundreds in the pores and side channels, we add the inertial term to the hydraulic resistance of the rectangular slit \cite{Jensen2014,Hadjaje2025}:
 \begin{equation}
\label{eqn:RslitHighRe} R_{slit}=\frac{32 \mu}{\pi a^2\,b} + \frac{\rho u_{slit}^2}{q_{slit}}
 \end{equation}
where $u_{slit}$ is the average velocity and $q_{slit}$ the flow rate through the slit, which simplifies in
 \begin{equation}
\label{eqn:RslitHighRe2} R_{slit}=\frac{32 \mu}{\pi a^2\,b} + \frac{\rho \, q_{slit}}{a^2 b^2}
 \end{equation}

We can define the hydraulic resistances of the two types of rectangular slits composing the first column, pores and side channels, assuming that $w_s<h$ and $\delta -d<h$:
\begin{eqnarray}
  \label{eqn:RPHighRe}  R_{P}&=&\frac{32 \mu}{\pi h(\delta-d)^2} + \frac{\rho \, q}{(n-1)\, h^2 (\delta-d)^2}\frac{q_{in}}{q}\\
\label{eqn:RSCHighRe} R_{SC}&=& \frac{32 \mu}{\pi h w_{s}^{2}} + \frac{\rho \, q}{2 \, h^2 w_s^2}\left(1-\frac{q_{in}}{q}\right),
\end{eqnarray}
The permeabilities of the pores and side channel are now functions of the flow rate $q$ or Reynolds number $Re$, and the flow rate entering the array $q_{in}$.

At intermediate Reynolds numbers, the drag on a cylinder is no longer purely viscous, and pressure drag becomes non-negligible. In Fig.\ref{fig:singlerow}(c), we present a map of the pressure field in the channel. The total pressure is rescaled with the inertial pressure $\rho U^2$. At low $Re$ the pressure in the channel is larger than the inertial component, due to viscous effects. At $Re = 1$, a low-pressure wake forms behind the outer cylinders. Eventually, as $Re$ increases, a low-pressure wake forms behind all the cylinders. The characteristic pressure in the wake of an isolated cylinder is of the order of $\rho U^2$ at intermediate $Re$  \cite{Thom1997}. In our system, the negative pressure in the wake of the cylinders is expected to increase the flow rate through the pore. 

To model the effects of inertia on the flow through and around a line of cylinders, we account for the dependence of the hydraulic resistance on the flow rate and the additional pressure drop in the pores due to the wake of the cylinders. We can therefore write the pressure drop in the side channels: 
\begin{equation}
\label{eqn:dPSCHighRe} dp =R_{SC} q_{SC} = R_{SC} \frac{q}{2} \left(1-\frac{q_{in}}{q}\right)   
\end{equation}
and in the pores,
\begin{equation}
\label{eqn:dPPHighRe} dp + \rho U^2 = R_{P} q_{P} = R_{P} \frac{q}{n-1}.\frac{q_{in}}{q} 
\end{equation}
Setting $dp$ equal, we get
\begin{equation}
\label{eqn:dPHighRe} \frac{R_{SC} \,q}{2} \left(1-\frac{q_{in}}{q}\right) + \rho \left(\frac{q}{hw}\right)^2 =  \frac{R_{P}\,q}{n-1} \frac{q_{in}}{q}  
\end{equation}

Using the definition of the hydraulic resistances, Eqs. \ref{eqn:RPHighRe} and \ref{eqn:RSCHighRe}, we obtain a quadratic equation for $q_{in}/q$:
\begin{multline}
\label{eqn:qinoverqHighRe} 
\frac{\rho q}{h^2}\left[ \left(\frac{1}{2w_s}\right)^2- \left(\frac{1}{(n-1)(\delta-d)}\right)^2 \right] \left(\frac{q_{in}}{q}\right)^2
- \left[ \frac{\rho q}{2h^2w_s^2}+\frac{32 \mu}{\pi h} \left(\frac{1}{2 w_s^2}+\frac{1}{(n-1)(\delta-d)^2}\right) \right] \frac{q_{in}}{q}
+ \\
\frac{16 \mu}{\pi h w_s^2}+\frac{\rho q}{h^2}\left(\frac{1}{4w_s^2}+\frac{1}{w^2}\right)= 0
\end{multline}
This equation admits one physical analytical solution, which is divided by $\alpha$ to obtain the leakiness $\Lambda$. The values of $\Lambda$, which now depend on $Re$, $C$, and $\phi$ are presented in Figs. \ref{fig:model}(b-d). We first compare the model with numerical results for one value of $Re$, $Re = 45$ in Fig. \ref{fig:model}(b). The results are in good agreement for the range of confinement and porosity values considered. Comparing the curves of identical color, corresponding to the same confinement in Figs. \ref{fig:model}(a-b), we note that the model captures the relative change in leakiness between the Stokes regime and $Re = 45$. 

We also compare the model with numerical results obtained for the range of Reynolds numbers, $Re \in [10^{-3}, 45]$ (see Figs. \ref{fig:model}(c-d)). At low $Re$, the model with the inertial correction yields the same values as the Sampson approximation. The model also captures the evolution of the leakiness with $Re$. The quantitative agreement, i. e., the difference between the values predicted by the model and the numerical simulations, increases with the porosity and confinement, reaching about 0.1 for $\phi = 0.63$ and $C = 0.89$. Yet the relative error is a decreasing function of the porosity and confinement, with values ranging from 6.4 for $\phi = 0.89$ and $C = 0.7$ to 22 $\%$ for $\phi = 0.63$ and $C = 0.38$, when the flow rate through the pores is low. At very high confinement for $C = 0.9$, the geometry of the flow in the small side channel differs significantly from the conditions under which the Sampson flow is derived, i.e., for an isolated orifice in an infinite plane. The no-slip condition at the wall leads to a non-symmetrical flow distribution about the vertical mid-plane of the slit, which is not accounted for in the first-order model. In addition, the model assumes that viscous dissipation in the pores and side channels is negligible, which is not the case in a narrow channel near a no-slip boundary. At intermediate $Re$, the low-pressure conditions behind the cylinders are expected to extend into the side channel. Both of those effects could contribute to an over-estimation of $\Lambda$. Yet for $C<0.9$, the simple model presented here predicts the values of $\Lambda$ for the two end values of $Re$, $Re = 10^{-3}$ and $Re = 45$, and the transition between the two values within the correct range of $Re$. This predictive model for a single column of cylinders can now be used to rationalize the flow regimes in an array.

\section{Discussion}\label{sec: Discussion}
The analytical model focuses on predicting the flow through a line of cylinders, as the flow rate entering a single column is comparable to that entering an array. The comparison of the model and the numerical simulations for a column and an array in Figs. \ref{fig:model}(a,c,d) shows that indeed the model for the leakiness of a single column captures reasonably well the leakiness of the array. The leakiness of the array is systematically slightly lower than the leakiness of the column, except at large confinement ($C=0.9$). However, the differences due to column interactions are not large enough to be captured by a first-order model.
Using the modeled values of the leakiness, we can predict the regime that the system will experience at low $Re$. Using Fig. \ref{fig:model}(a), we see that the rake regime is expected for a decreasing range of porosity as the confinement increases. At $C= 0.1$, all porosities are in the rake regime. At $C = 0.9$, systems with porosity $\phi>0.5$ will be in the sieve regime. The threshold porosity for the rake regime as a function of the array confinement can be obtained analytically by setting $\Lambda = 0.1$, in Eq. \ref{eqn:LambdaLowReHighC}. In addition, the inertial corrections allow predicting the maximum $Re$ at which a system is in the rake regime. Fig. \ref{fig:model}(d) illustrates this result for an array of porosity $\phi = 0.63$ and low confinement $C = 0.38$. The system is expected to leave the rake regime $Re \approx 1.6$. This is consistent with the numerical results, which show a transition between $Re = 1$ and $Re = 2$.
Because the model is limited to the first row of the array, it can predict the flow regime at low $Re$ and the amount of fluid entering the array as a function of porosity and confinement over the range of $Re$ considered. Although a more advanced model would be needed to predict whether the array is in the deflection or sieve regime for $\Lambda > 0.1$, we note that most of the fluid exiting the array through the side tends to leave the array between the first and second column of cylinders. We can therefore use the model results to compare the flow conditions in and around the array between the first and second columns. The model predicts the flow through the array $q_{in}$ as a function of $C$, $\phi$, and $Re$. We can use mass conservation to obtain $q_{side}$ in the side channel. Knowing the flow rates, we can estimate the velocity in the pore and side channel and define the inertial pressure difference between a pore and a side channel:
\begin{equation}
dp = \rho \left(u_{side}^2 - u_{pore}^2 \right).
\end{equation}
When the velocity in the side channel exceeds that in the pore, the pressure difference favors flow from the pore to the side channel, referred to as deflection flow. The deflection flow is resisted by viscous dissipation at low $Re$, $Re<1$, and inertial stresses at $Re > 1$. We therefore estimate a viscous and an inertial stress based on the fluid velocity in the pore: $\tau_v = \mu u_{pore}/d$ and $\tau_i = \rho u_{pore}^2$. The largest of those stresses limits the side flow. We compare the inertial pressure with the viscous stress at low $Re$, and with the inertial stress at larger $Re$. The threshold value for $Re$, above which inertial effects dominate, is when $\tau_v = \tau_i$, typically around $Re=1$. The results are presented in Figs. \ref{fig:model}(e-f) with solid lines correspond to the ratio of the inertial pressure to the largest resisting stress, the dashed line are the ratio of the inertial pressure to the lower, negligible resisting stress. The results show that the inertial pressure is favorable to flow exiting the array until large $Re$, except at $C=0.9$, where the flow is highly confined. This is actually consistent with anecdotal experimental observations of fluid entering the array through the side at high confinement. Although the pressure difference is favorable, the viscous stresses prevent deflection flow at low $Re$, which is consistent with a flow angle $\theta = 0^o$ at low $Re$ (see Figs. \ref{fig:num}(e-h). Similarly, as the fluid inertia becomes large at $Re = 40$, the ratio of the inertial pressure to the inertial stresses becomes small and negative. At intermediate values of $Re$, the ratio of the inertial pressure over the resisting stress increases as the viscous effects dominate, reaches a maximum when inertial and viscous effects are comparable, and decreases as the inertial effects become larger. This increase and decrease in rescaled inertial pressure as $Re$ increases is consistent with the increase and decrease in flow angle over the same range of $Re$. The maximum rescaled inertial pressure shows the same dependence on porosity and confinement as the maximum flow angle. Low values of the maximum rescaled inertial pressure correspond to the sieve regime, whereas high values correspond to the deflection regime. Even though the model is based on the first line of cylinders, the flow rates through the first pores and side channels are consistent with the numerical results and provide a rationale for the transport of fluid out of the array and into the side channels.

 \section{Conclusion}\label{sec: Conclusion}
This study combines experiments, numerical simulations, and analytical modeling to characterize and rationalize the influence of confinement on creeping flows through cylinder arrays. We show that the flow through the array strongly depends on the Reynolds number of the flow, the confinement  and porosity of the array. We describe the flow through the array with two parameters, the leakiness and the flow angle. At large porosity and confinement, the relative flow rate through the array is larger than $0.1$, and the flow angle remains below $45^o$ as the fluid is focused into the array. At low confinement and porosity, the fluid flows around the array, so the leakiness is low, and the flow angle is large. The leakiness and flow angle define the flow regime in the array. At low $Re$, the arrays are in the rake regime for low confinement or in the sieve regime for high confinement. As $Re$ increases, the leakiness exceeds 0.1 for all arrays, so they are in the sieve or deflection regimes at $Re \approx 40$. We develop a first-order model to predict the leakiness of a square array of n by n cylinders and demonstrate its validity for $n = 5$. The approach significantly simplifies the system by considering only a line of cylinders and modeling them as a membrane with rectangular holes. Yet the model predicts the leakiness of the array and rationalizes the deflection flow for a confined array of cylinders.
In closing, the results show the complex dependence of permeability on porosity, confinement, and the flow Reynolds number. The model of the transport through the first line of cylinders should be a first step toward a more advanced description of the full array and the design of porous media with the specific flow properties for applications ranging from particle filters to lightweight wings. For such applications, future work should address hydrodynamic loading as a function of flow regime.

\begin{acknowledgments}
E.D. and S.K.B. acknowledge partial support from the Army
Research Oﬃce under grant no. $W911NF-23-2-0046$. N.D.J. acknowledges support from the Edison Program at UCSB. J.P.R's work was partially supported by the MRSEC Program of the National Science Foundation under Award No. DMR 2308708 through the FLAM Summer Internship Program.
\end{acknowledgments}

All authors were involved in the design of the project. S.S.T, N.D.J, J.P.R and S.K.B conducted the experiments and simulations. All authors analyzed the data, interpreted the results, developed the model, and contributed to the manuscript. E.D. conceptualized and guided the project

The authors declare no competing interests.

The data that support the findings of this article are openly available at [temporary \href{https://datadryad.org/share/LINK_NOT_FOR_PUBLICATION/3OiMzrIGUEERNvg21l9VcAxnDXwhVvqgmVNsv6lVU7k}{link}, data awaiting curation].
\bibliography{export_2026-2-24}
\end{document}